\documentclass[10pt,twoside]{article}
\usepackage{graphicx}
\usepackage{amsmath,amssymb}
\def\hang{\hangindent\parindent}
\def\textindent#1{\indent\llap{#1\enspace}\ignorespaces}

\begin{document}

\thispagestyle{plain}

\centerline{\Large\bf  Synchronization in delayed discrete-time
complex networks\footnote[1]{\footnotesize The current work was
supported in part by the Tianyuan Foundation (A0324651).} }

\vspace{0.5cm}

\centerline{Weigang Sun$^a$, Changpin
Li$^{a,\,}$\footnote[2]{\footnotesize Corresponding author.
Department of Mathematics, Shanghai University, Shanghai 200444.
Tel: 86-21-66133906; Fax: 86-21-66133292.
 {E-mail address:} leecp@online.sh.cn} and Zhengping Fan$^b$}

\vskip 2pt \centerline{\footnotesize\it $^a$ Department of
Mathematics,~Shanghai University, Shanghai 200444, China}

\centerline{\footnotesize\it $^b$ Department of Electronic
Engineering, City University of Hong Kong, Hong Kong, China}

\vspace{0.5cm}

\noindent{\bf Abstract}

{\small In this paper, we study synchronization in the delayed
discrete-time complex networks. Several criterions of
synchronization stability for such networks are established. And
illustrative examples are presented. The numerical simulations
coincide with the theoretical analysis.}
\vskip 5pt

\noindent{{\sl PACS:} 05.45.Ra, 05.45.Xt}

\noindent{{\sl  Keywords:}  {Discrete-time complex network, time
delay, synchronization }}

\bigbreak \noindent{\bf 1. Introduction}
\medskip

Time delays commonly exist in the real networks, some of them are
trivial so can be ignorant whilst some of can not be ignored, such
as in the long-distance communication and traffic congestions,
etc. So such networks with retard time attracts much attention.
Recently, Li, {\sl et al.}, studied the following continuous
network model with time delays [1],
$$
\dot{x}_i=f(x_i) +\varepsilon\sum_{j=1}^Nc_{ij}
A(x_{j1}(t-\tau_1),x_{j2}(t-\tau_2),\cdots,x_{jn}(t-\tau_n))^T\triangleq
f(x_i) +\varepsilon\sum_{j=1}^Nc_{ij} A\cdot
\overline{x_j(t-\tau)}\,, \eqno (1)
$$
where $f:\Re^n\rightarrow\Re^n $ is a continuously differentiable
function, $x_i=(x_{i1},x_{i2},\cdots,x_{in})^T\in\Re^n $ are the
state variables of node $i$,  $\varepsilon>0$ represents the
coupling strength, $A=(a_{ij})_{n\times n}\in\Re^{n\times n}$
indicates inner-coupling between the elements of the node itself,
while $C=(c_{ij})_{N\times N}$ denotes the outer-coupling between
the nodes of the whole network (it is often assumed that there is
at most one connection between node $i$ and another node $j$, and
that there are no isolated clusters, i.e., $C$ is an irreducible
matrix). The entries $c_{ij }$ are defined as follows: if there is
a connection between node $i$ and node $j\,(j\neq i)$, then we set
$c_{ij}=1$; otherwise $c_{ij}=0\,(j\neq i)$, and the diagonal
elements of $C$ are defined by $c_{ii}=-\sum_{j=1,j\neq i}^N
c_{ij},\,i=1,2,\cdots,N,$ $\tau_i,\,i=1,\cdots,n,$ are the time
delays.

In this paper, we study the discrete version of network (1), which
is described as,
$$
\begin{array}{rcl}{x}_i(k+1)&=&f(x_i(k)) +\varepsilon\sum_{j=1}^Nc_{ij}
A(x_{j1}(k-\tau_1),x_{j2}(k-\tau_2),\cdots,x_{jn}(k-\tau_n))^T\\
&\triangleq& f(x_i(k))
+\varepsilon\sum_{j=1}^Nc_{ij}A\cdot\overline{x_j(k-\tau)}\,,
\end{array}
\eqno (2)
$$
in which $f$, $\varepsilon$, $x_i~(i=1,2,\cdots,N)$, $C$ and $A$
have the same meanings as those in (1). The only difference is
that in Eq. (2), $x_i$ and $\tau_i$ are defined in the positive
integer set $\mathcal{Z^+}$.

In model (2), the inner-coupling is linear since $A$ is a matrix,
a natural generalization is that the inner-coupling can be
nonlinear [2]. Such a discrete-time network with retard time reads
as,
$$
\begin{array}{rcl}{x}_i(k+1)&=&f(x_i(k)) +\varepsilon\sum_{j=1}^Nc_{ij}
Ag(x_{j1}(k-\tau_1),x_{j2}(k-\tau_2),\cdots,x_{jn}(k-\tau_n))^T\\
&\triangleq& f(x_i(k))
+\varepsilon\sum_{j=1}^Nc_{ij}Ag(\overline{x_j(k-\tau)})\,,
\end{array}
 \eqno (3)
$$
where $f$, $\varepsilon$, $x_i,~\tau_i~(i=1,2,\cdots,N)$ and $C$
have the same meanings as those in (2), $g:\Re^n\rightarrow\Re^n $
is a continuously differentiable nonlinear function.

The outer-coupling configuration $C$ in networks (2) and (3) has
following properties [3,4].

\medskip
\textsl{Lemma 1:} Suppose that $C=(c_{ij})_{N\times N}$ is a real
symmetric and irreducible matrix, where $c_{ij}\ge 0~(i\ne
j),\,c_{ii}=-\sum_{j=1,j\ne i}^N c_{ij}$, then

(1) 0 is an eigenvalue of $C$ with multiplicity 1, associated with
eigenvector $(1,1,\cdots,1)^T$;

(2) all the other eigenvalues of $C$ are less than 0;

(3) there exists a unitary matrix,
$\Phi=(\phi_1,\phi_2,\cdots,\phi_N)$ such that
$$
C^T\phi_k=\lambda_k\phi_k,~~k=1,2,\cdots, N,
$$
where $0=\lambda_1>\lambda_2\ge\lambda_3\ge\cdots\ge\lambda_N$ are
the eigenvalues of $C$.

\medskip
In what follows, the definition of synchronization for networks
(2) and (3) is introduced below [2].

\medskip
\textsl{Definition 1:} Let $\mathcal{A}$ be an attractor of the
discrete-time dynamical system
$$
s(k+1)=f(s(k)).
$$
We say that networks (2) and (3) are (asymptotically) synchronized
to $\mathcal{A}$, if for $k\longrightarrow +\infty$,
$$
x_i\longrightarrow\mathcal{A},\,i=1,\,\cdots,\,N.
$$

In the rest of this paper, the criterions of synchronization
stability for networks (2) and (3) are established in Section 2.
And the numerical examples are presented in Section 3.

\medskip

\noindent{\bf 2. Synchronization theorems}

\medskip
By utilizing Lemma 1, one can derive the following theorem.

\medskip

\textsl{Theorem 1:} Consider the delayed discrete-time network
(2), let $0=\lambda_1>\lambda_2\ge \cdots\ge\lambda_N $ be the
eigenvalues of the coupling configuration matrix $C$. If the
following $N-1$ systems of $n$-dimensional linear delayed
equations are asymptotically stable about their zero solutions,
$$
 \eta(k+1)=Df(s(k))\eta(k)+\varepsilon\lambda_i A\cdot
 \overline{\eta(k-\tau)}\,,~~i=2,\cdots, N,\eqno(4)
$$
where $Df(s(k))\in \Re^{n\times n}$ is the Jacobian of $f(x(k))$
at $s(k)$, $\eta(k)\in \Re^n$,
$\overline{\eta(k-\tau)}=(\eta_1(k-\tau_1),\cdots,\eta_n(k-\tau_n))^T\in
\Re^n$, $s(k)$ is the orbit of attractor $\mathcal{A}$ of equation
$s(k+1)=f(s(k))$, then network (2) is synchronized to the
attractor $\mathcal{A}$.

\textsl{Proof:} Linearizing (2) at $s(k)$ yields,
 $$
e_i(k+1)=Df(s(k)e_i(k))+\varepsilon\sum_{j=1}^Nc_{ij} A\cdot
\overline{e_j(k-\tau)},~1\leq i\leq N,
$$
where $e_i(k)$ denotes the deviation from the state $s(k)$, i.e.,
$ e_i(k)=x_i(k)-s(k)$,
$\overline{e_j(k-\tau)}=(e_{j1}(k-\tau_1),\cdots,e_{jn}(k-\tau_n))^T\in
\Re^n$. This linearized system can be rewritten as
$$
e_i(k+1)=Df(s(k))e_i(k)+\varepsilon
A\cdot(\overline{e_1(k-\tau)}\,,\overline{e_2(k-\tau)}\,,\cdots
,\overline{e_N(k-\tau)}\,)(c_{i1},\cdots, c_{iN})^T.
$$
Let $ e(k)=(e_1(k),e_2(k),\cdots,e_N(k))\in\Re^{n\times N}$, the
above equation can be expressed in a compact form,
$$
e(k+1)=Df(s(k))e(k)+\varepsilon A\cdot \overline{e(k-\tau)}\,C^T,
$$
where
$\overline{e(k-\tau)}=(\overline{e_1(k-\tau)}\,,\overline{e_2(k-\tau)}\,,\cdots
,\overline{e_N(k-\tau)}\,)\in \Re^{n\times N}$.

From Lemma 1, there exists a nonsingular matrix $\Phi$, such that
$C^T\Phi=\Phi\Gamma,\,\Gamma={\rm
diag}(\lambda_1,\cdots,\lambda_N)$. If one sets
$e(k)\Phi=v(k)=(v_1(k),v_2(k),\cdots,v_N(k))\in\Re^{n\times N}$,
then the above equation can be transformed into the following
matrix equation
$$
v(k+1)=Df(s(k))v(k)+\varepsilon A\, \overline{v(k-\tau)}\,\Gamma,
$$
which immediately follows that,
$$
v_i(k+1)=Df(s(k))v_i(k)+\varepsilon\lambda_i A
\cdot\overline{v_i(k-\tau)}\,,~i=1,\cdots,N,
$$
where
$\overline{v_i(k-\tau)}=(v_{i1}(k-\tau_1),\cdots,v_{in}(k-\tau_n))^T\in
\Re^n$.

Note that $\lambda_1=0$ corresponds to the linearized system of
the isolate equation $s(k+1)=f(s(k))$. If the following $N-1$
pieces of $n$-dimensional linear time-varying delayed equations
$$
v_i(k+1)=Df(s(k))v_i(k)+\varepsilon\lambda_i A\cdot
\overline{v_i(k-\tau)},~i=2,\cdots,N,
$$
are asymptotically stable around their zero solutions, then $e(k)$
will tend to zero, which shows that the conclusion holds. This
completes the proof.
\medskip

\textsl{Theorem 2:} Assume that all eigenvalues of the matrix $C$
in (2) are listed in order, $ 0=\lambda_1>\lambda_2\geq\cdots
\geq\lambda_N.$ If there exists a positive-definite matrix $P>0$
such that
$$
\left[%
\begin{array}{cc}
  Df(s(k))^TPDf(s(k))-P+I & \varepsilon \lambda_iDf(s(k))^TPA \\
   \varepsilon \lambda_iA^TPDf(s(k))& \varepsilon^2\lambda_i^2A^TPA-I \\
\end{array}%
\right]<0,\,\, i=2,3,\cdots,N,\eqno(5)
$$
where $I$ is the identity matrix, $s(k)$ is the orbit of
attractor $\mathcal{A}$ of the equation $s(k+1)=f(s(k))$, then
network (2) is synchronized to $\mathcal{A}$ for any fixed delay
$\tau_i\in \mathcal{Z^+}\,~(i=1,2,\cdots,n)$.

\textsl{Proof:} Select a Lyapunov functional
as
$$
V(\eta(k))=\eta(k)^TP\eta(k)+\sum_{i=1}^n
\sum_{\sigma=k-\tau_i}^{k-1}\eta_i(\sigma)^T\eta_i(\sigma),
$$
in which $\eta(k)=(\eta_1(k),\eta_2(k),\cdots,\eta_n(k))^T$. Then,
along the solution of the $i$th ($i=2,3,\cdots,N$) equation in
system (4), one gets
$$
\begin{array}{rcl} \triangle V(\eta(k))
&=& V(\eta(k+1))- V(\eta(k))\\
&=& [Df(s(k))\eta(k)+\varepsilon\lambda_i A\cdot
 \overline{\eta(k-\tau)}]^TP[Df(s(k))\eta(k)+\varepsilon\lambda_i A\cdot
 \overline{\eta(k-\tau)}]\\
 &&-\eta(k)^TP\eta(k)+\eta(k)^T\eta(k)-\overline{\eta(k-\tau)}^T\cdot\overline{\eta(k-\tau)}\\
&=&\left[\begin{array}{c}
  \eta(k) \\
  \overline{\eta(k-\tau)} \\
  \end{array}\right]^T
  \left[%
\begin{array}{cc}
  Df(s(k))^TPDf(s(k))-P+I & \varepsilon \lambda_iDf(s(k))^TPA \\
   \varepsilon \lambda_iA^TPDf(s(k))& \varepsilon^2\lambda_i^2A^TPA-I\\
\end{array}%
\right] \left[\begin{array}{c}
  \eta(k) \\
  \overline{\eta(k-\tau)} \\
\end{array}\right]
\end{array}
$$

By using linear matrix inequality (LMI) (5), one has $\triangle
V(\eta(k))<0$ for all $i=2,3,\cdots,N$, which implies that the
zero solutions of systems (4) are asymptotically stable. From
Theorem 1, network (2) is synchronized to $\mathcal{A}$. The proof
is finished.
\medskip

In the following, one can similarly obtain theorems 3 and 4 for
network (3).
\medskip

\textsl{Theorem 3:} Consider the delayed discrete-time network
(3), let $ 0=\lambda_1>\lambda_2\ge \cdots\ge\lambda_N$ be the
eigenvalues of the coupling configuration matrix $C$. If the
following $N-1$ systems of $n$-dimensional linear delayed
equations are asymptotically stable about their zero solutions,
$$
 \eta(k+1)=Df(s(k))\eta(k)+\varepsilon\lambda_i ADg(\overline{s(k-\tau)})\cdot
 \overline{\eta(k-\tau)}\,,~~i=2,\cdots, N,
$$
where $Df(s(k)),Dg(\overline{s(k-\tau)})\in \Re^{n\times n}$ are
the Jacobians of $f(x(k)),g(x(k))$ at $s(k)$ and
$\overline{s(k-\tau)}$, $\eta(k)\in \Re^n$,
$\overline{s(k-\tau)}=(s_1(k-\tau_1),\cdots,s_n(k-\tau_n))^T\in\Re^n$,
$\overline{\eta(k-\tau)}=(\eta_1(k-\tau_1),\cdots,\eta_n(k-\tau_n))^T\in
\Re^n$, $s(k)$ is the orbit of attractor $\mathcal{A}$ of the
equation $s(k+1)=f(s(k))$, then network (3) is synchronized to
$\mathcal{A}$.

\medskip
\textsl{Theorem 4:} Assume that all eigenvalues of the matrix $C$
in (3) are listed in order, $ 0=\lambda_1>\lambda_2\geq\cdots
\geq\lambda_N.$ If there exists a positive-definite matrix $P>0$
such that
$$
\left[%
\begin{array}{cc}
  Df(s(k))^TPDf(s(k))-P+I & \varepsilon \lambda_iDf(s(k))^TPADg(\overline{s(k-\tau)}) \\
 \varepsilon \lambda_iDg(\overline{s(k-\tau)})^TA^TPDf(s(k))& \varepsilon^2\lambda_i^2
 Dg(\overline{s(k-\tau)})^TA^TPADg(\overline{s(k-\tau)})-I \\
\end{array}%
\right]<0,\,\,i=2,\cdots,N,
$$
where $I$ is the identity matrix, $s(k)$ is the orbit of attractor
$\mathcal{A}$ of equation $s(k+1)=f(s(k))$, then network (3) is
synchronized to $\mathcal{A}$ for any fixed delay $\tau_i\in
\mathcal{Z^+}\,~(i=1,2,\cdots,n)$.

\medskip

In [1], by using ``matrix measure" [5,6], Li, {\sl et al.},
discussed the synchronization of network (1) and the following
network (6),
$$
\dot{x}_i=f(x_i) +\varepsilon\sum_{j=1}^Nc_{ij}
A(x_{j1}(t-\tau_{j1}),x_{j2}(t-\tau_{j2}),\cdots,x_{jn}(t-\tau_{jn}))^T\triangleq
f(x_i) +\varepsilon\sum_{j=1}^Nc_{ij} A\cdot
\overline{x_j(t-\tau_j)}\,, \eqno (6)
$$
where $f$, $\varepsilon$, $x_i~(i=1,2,\cdots,N)$, $C$ and $A$ have
the same meanings as those in (1), the sole difference is that in
(6) a different node $j$ has a different time-delay vector,
$(\tau_{j1},\tau_{j2},\cdots,\tau_{jn})$.

We can also apply ``LMIs" presented here to establishing
synchronization theorems for  delayed continuous-time networks [1]
and [6]. In the following, the definition of synchronization is
given.

\medskip
\textsl{Definition 2} [1,2]: Assume that $\mathcal{B}$ is an
attractor of the continuous dynamical system
$$
\dot{s}=f(s).
$$
We say that networks (1) and (6) are (asymptotically) synchronized
to $\mathcal{B}$, if for $t\longrightarrow +\infty$,
$$
x_i\longrightarrow\mathcal{B},\,i=1,\,\cdots,\,N.
$$
\medskip

Here, we list the synchronization theorems of networks (1) and (6)
just for reference later on. For details of the proofs, see [7].

\medskip
\textsl{Theorem 5:} Consider the delayed continuous network (1),
let $0=\lambda_1>\lambda_2\ge \cdots\ge\lambda_N$ be the
eigenvalues of the coupling configuration matrix $C$. If the
following $N-1$ systems of $n$-dimensional linear time-varying
delayed differential equations are asymptotically stable about
their zero solutions,
$$
 \dot\eta(t)=Df(s(t))\eta(t)+\varepsilon\lambda_i A\cdot
 \overline{\eta(t-\tau)}\,,~~i=2,\cdots, N,
$$
where $Df(s(t))\in \Re^{n\times n}$ is the Jacobian of $f(x(t))$
at $s(t)$, $\eta(t)\in \Re^n$,
$\overline{\eta(t-\tau)}=(\eta_1(t-\tau_1),\cdots,\eta_n(t-\tau_n))^T\in
\Re^n$, $s(t)$ is an orbit of attractor $\mathcal{B}$ of equation
$\dot{s}=f(s)$, then network (1) is synchronized to $\mathcal{B}$.
\medskip

\textsl{Theorem 6:} Assume that all eigenvalues of the matrix $C$
in (1) are listed in order, $0=\lambda_1>\lambda_2\geq\cdots
\geq\lambda_N$. If  there exists a positive-definite matrix $P>0$
such that
$$
\left[%
\begin{array}{cc}
  PDf(s(t))+Df(s(t))^TP+I & \varepsilon \lambda_iPA \\
   \varepsilon \lambda_iA^TP& -I \\
\end{array}%
\right]<0,\,\,i=2,3,\cdots,N,
$$
where $I$ is the identity matrix, $s(t)$ is an orbit of the
attractor $\mathcal{B}$ of the equation $\dot{s}=f(s)$, then
network (1) is synchronized to $\mathcal{B}$ for any fixed delay
$\tau_k>0\,(k=1,2,\cdots,n)$.
\medskip

\textsl{Theorem 7:} Consider the delayed dynamical network (6),
let $0=\lambda_1>\lambda_2\ge \cdots\ge\lambda_N$ be the
eigenvalues of the coupling configuration matrix $C$. If the
following $N-1$ systems of $n$-dimensional linear time-varying
delayed differential equations are asymptotically stable about
their zero solutions,
$$
 \dot\eta(t)=Df(s(t))\eta(t)+\varepsilon\lambda_i A\cdot
 \overline{\eta(t-\tau_i)}\,,~~i=2,\cdots, N,
$$
where $Df(s(t))\in \Re^{n\times n}$ is the Jacobian of $f(x(t))$
at $s(t)$, $\eta(t)\in \Re^n$,
$\overline{\eta(t-\tau_i)}=(\eta_1(t-\tau_{i1}),\cdots,\eta_n(t-\tau_{in}))^T\in
\Re^n$, $s(t)$ is the stable equilibrium $\mathcal{B}_0$ of
equation $\dot{s}=f(s)$, then network (6) is synchronized to
$\mathcal{B}_0$.
\medskip

\textsl{Theorem 8:} Assume that all eigenvalues of the matrix $C$
in (6) are listed in order, $0=\lambda_1>\lambda_2\geq\cdots
\geq\lambda_N$. If there exists a positive-definite matrix $P>0$
such that
$$
\left[%
\begin{array}{cc}
PDf(s(t))+Df(s(t))^TP+I & \varepsilon \lambda_iPA \\
\varepsilon \lambda_iA^TP& -I \\
\end{array}%
\right]<0,\,\,i=2,3,\cdots,N,
$$
where $I$ is the identity matrix, $s(t)$ is the equilibrium
$\mathcal{B}_0$ of equation $\dot{s}=f(s)$, then network (6) is
synchronized to $\mathcal{B}_0$ for any fixed delay $\tau_{kl}
>0\,(k,l=1,2,\cdots,n)$.
\medskip

\textsl{Remark 1:} Obviously, $\mathcal{B}_0
\subsetneq\mathcal{B}$. $\mathcal{B}$ also contains other stable
limit sets, e.g., stable (quasi-)periodic orbit, strange
attractor. Theorems 7 and 8 hold only for $\mathcal{B}_0$. For the
other cases, the studies are not easy but need further
consideration [8].

\bigbreak \noindent{\bf 3. Illustrative examples}
\medskip

In this section, we consider a five-node network, in which each
node is a simple 2-dimensional H\'enon map [9,10], described by
$$
 \left[\begin{array}{c}
 f_1(x_1,x_2)\\
f_2(x_1,x_2) \\
 \end{array}\right]=\left[\begin{array}{c}
   1+x_2-ax_1^2 \\
   bx_1 \\
 \end{array}\right] \eqno (6)
$$
If $a=0.5,\,b=0.3$, (6) has a period solution [9].

We use the coupled configuration matrix $(c_{ij})_{N\times N}$
[11], which is
$$
C= \left[%
\begin{array}{ccccc}
  -2 & 1 & 0 & 0 & 1 \\
  1 & -3 & 1 & 1 & 0 \\
  0 & 1 & -2 & 1 & 0 \\
  0 & 1 & 1 & -3 & 1 \\
  1 & 0 & 0 & 1 & -2 \\
\end{array}%
\right]
$$
whose eigenvalues are
$\lambda_1=0,\,\lambda_2=-1.382,\,\lambda_3=-2.382,\,\lambda_4=-3.168,\,
\lambda_4=-4.168$, and  use the inner-coupling matrix $A={\rm
diag}(1,1)$.

At first, the time-delay vector $(\tau_1,\,\tau_2)=(1,\,2)$ is
considered.

Here, set the coupling strength $\varepsilon=0.015$. By using the
MATLAB LMI Toolbox, one can obtain that there exists a common
positive-definite matrix,
$$
P=\left[%
\begin{array}{cc}
  5.3626 & 0 \\
  0 & 9.1633 \\
\end{array}%
\right]
$$
such that the condition of Theorem 2 is satisfied. From Theorem 2,
we know that this delayed network is synchronized to the stable
periodic state of the isolated H\'enon map. The following
numerical simulations are also in line with the above theoretical
analysis.

In Fig 1, we plot the curves of the synchronization errors between
the states of node $i$ and the $1st$ node, that is,
$e_{ij}=x_{ij}(k)-x_{1j}(k)$ for $i=2,\cdots,5, j=1,2$.

\begin{figure}[htb]
  {a)}{\includegraphics[width=7cm]{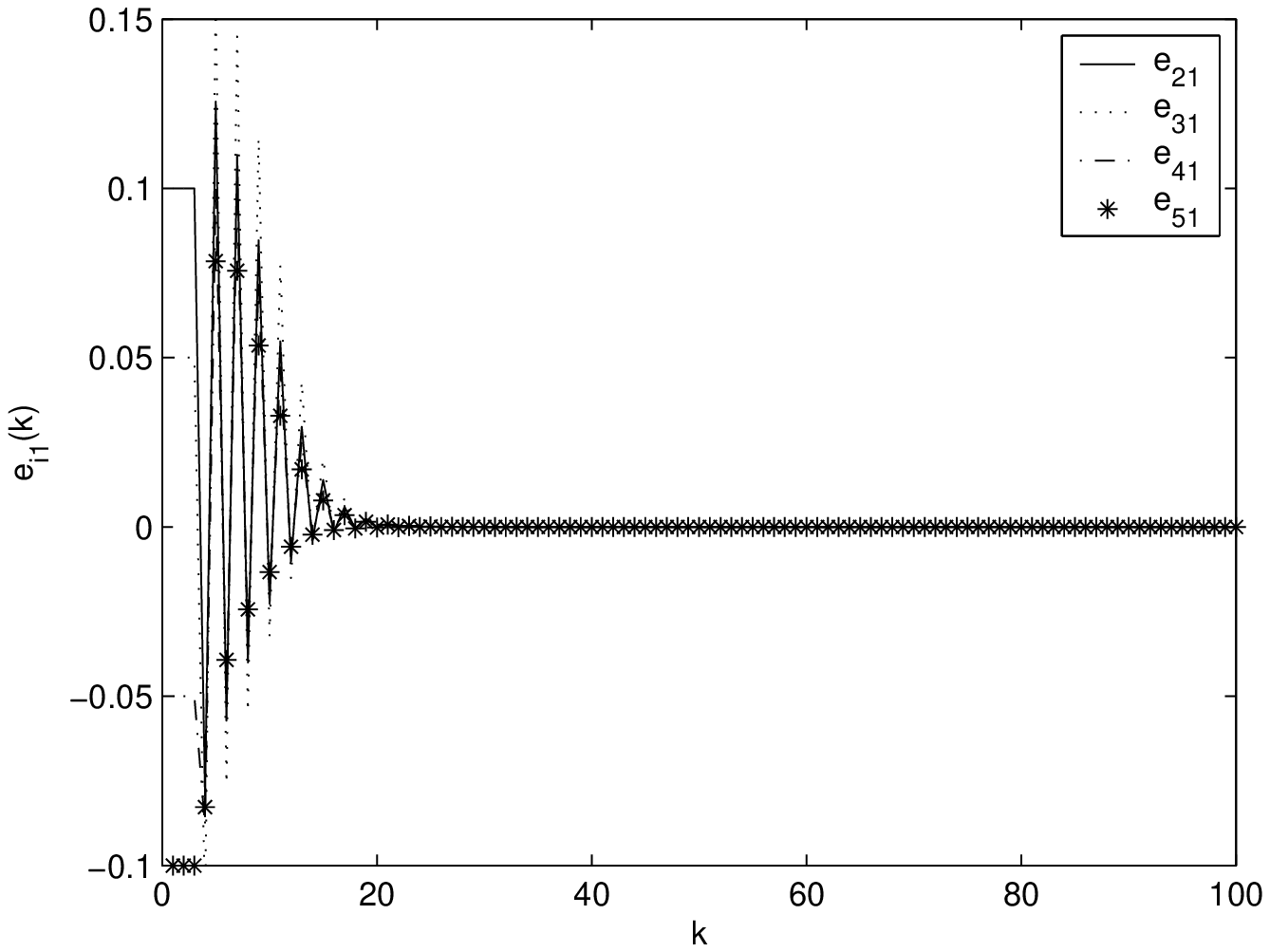}\hfill
  {b)}\includegraphics[width=7cm]{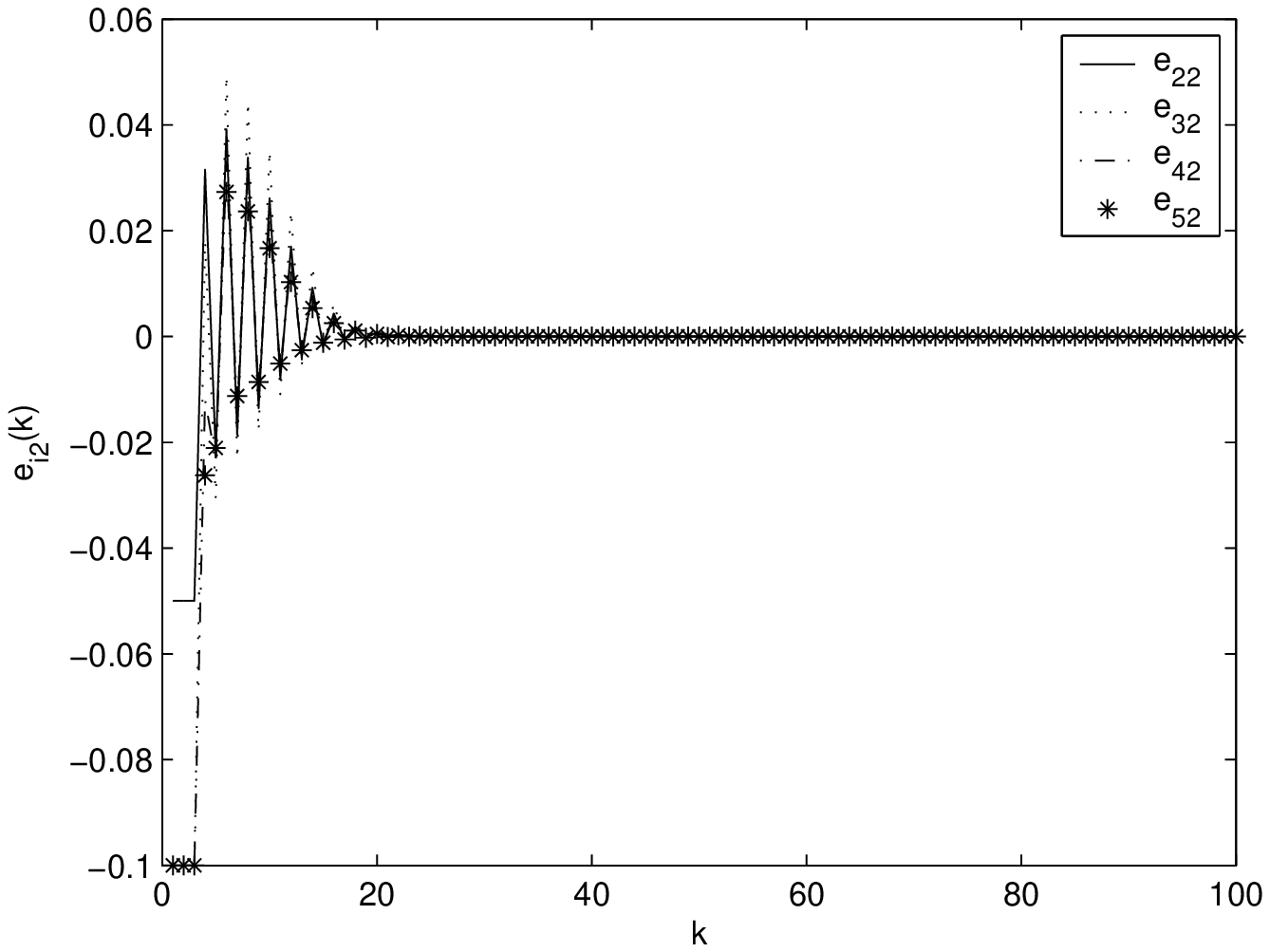}}
\end{figure}

\vskip -10pt{{\footnotesize Fig 1. Synchronization errors for the
coupled network with time-delay vector
$(\tau_1,\,\tau_2)=(1,\,2)$. (a) $j=1$ (b) $j=2$.}}

\medskip
Secondly, we choose the nonlinear inner-coupling function as
$$
\left[\begin{array}{c}
g_1(x_1,x_2)\\
g_2(x_1,x_2) \\
\end{array}\right]=\left[\begin{array}{c}
 e^{x_1} \\
 \sin\,x_2 \\
 \end{array}\right].\eqno (7)
$$

By using the MATLAB LMI Toolbox again, we can find there exists a
positive-definite matrix
$$
P=\left[%
\begin{array}{cc}
  5.3649 & 0 \\
  0 & 9.1714 \\
\end{array}%
\right]
$$
such that the condition of Theorem 4 is satisfied. So the
five-node H\'enon network with time delay
$(\tau_1,\,\tau_2)=(1,\,2)$ is synchronized to the stable periodic
state of the isolated H\'enon map. The numerical simulation below
shows this point of view.

In Fig 2, we plot the curves of the synchronization errors between
the states of node $i$ and node 1, i.e.,
$e_{ij}=x_{ij}(k)-x_{1j}(k)$ for $i=2,\cdots,5,\, j=1,2$.

\begin{figure}[htb]
  {a)}{\includegraphics[width=7cm]{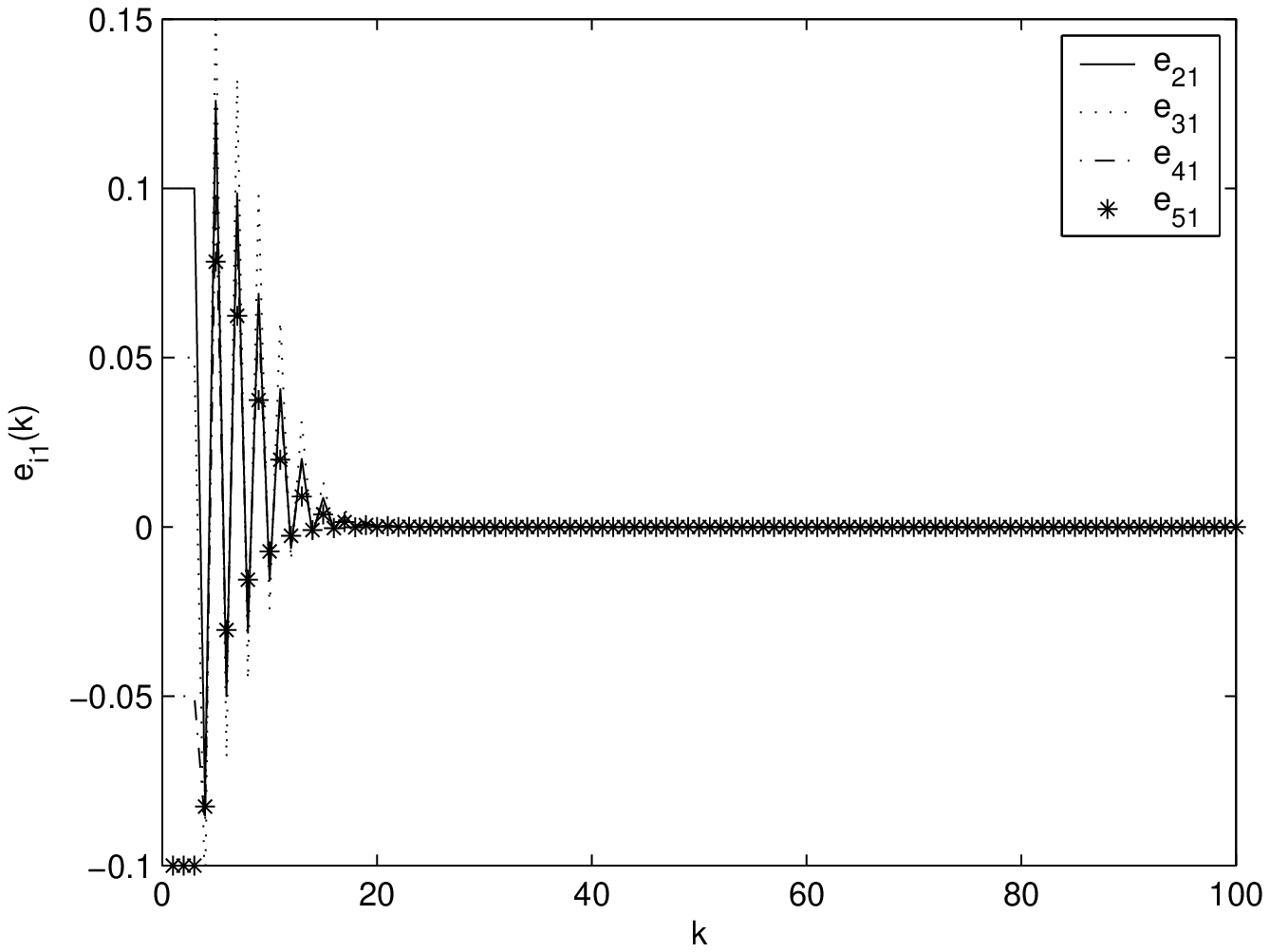}\hfill
  {b)}\includegraphics[width=7cm]{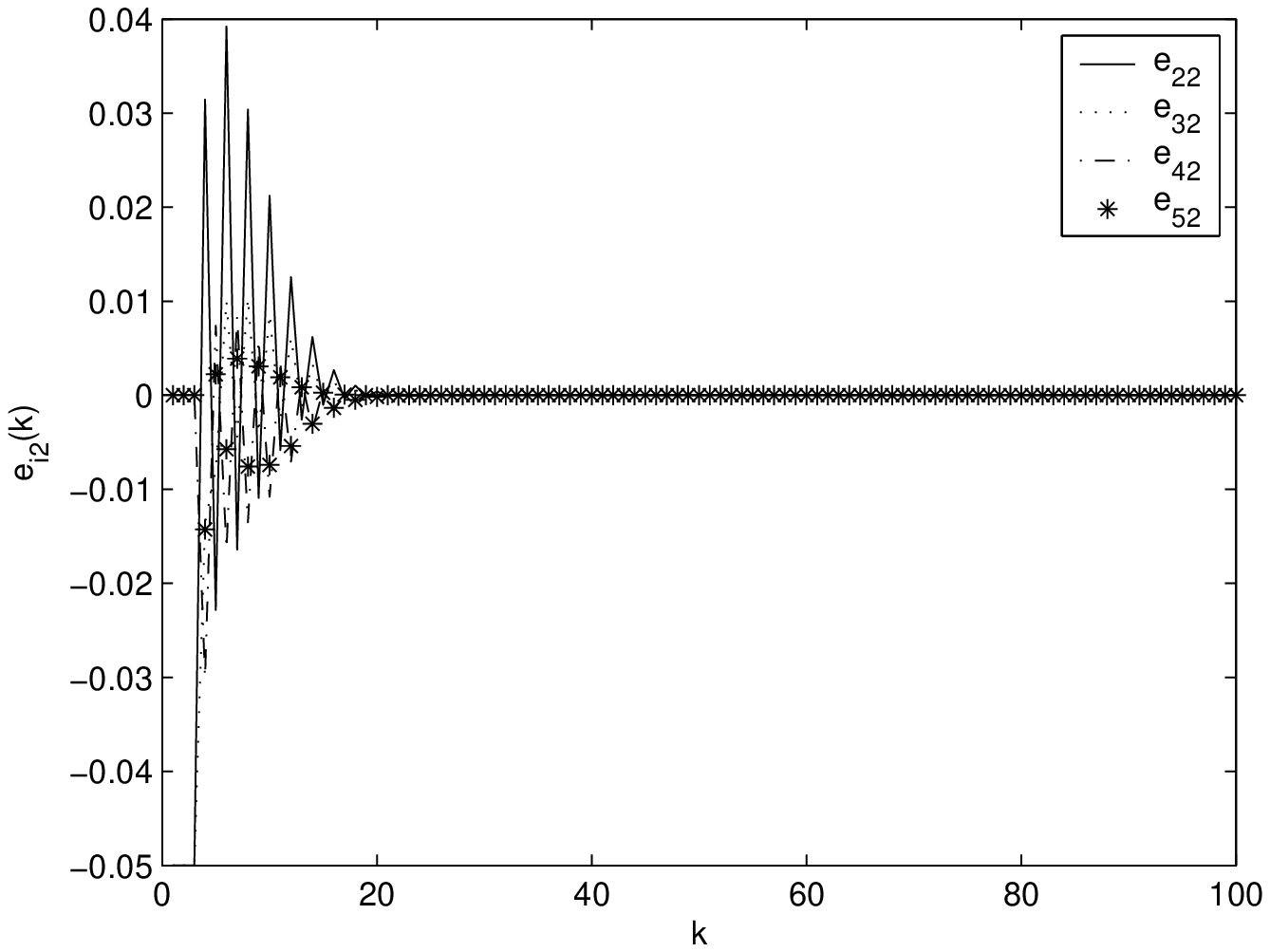}}
\end{figure}

\vskip -10pt{\centerline {\footnotesize Fig 2. Synchronization
errors for the five-node H\'enon network. (a) $j=1$ (b) $j=2$.}}

\bigbreak
 \bigbreak
\noindent

\def\item{\par\hang\textindent}
 \noindent{\small {\bf References}
\medskip

\item{[1]} C. P. Li, W. G. Sun and J. Kurths, ``Synchronization of
complex dynamical networks with time delays," accepted for
publication in Physica A.

\item{[2]} C. P. Li, G. Chen and T. S. Zhou, ``Some remarks on
synchronization of complex networks with nonlinear inner-coupling
functions," The 2nd Chinese Academic Forum on complex Dynamical
Networks, Beijing, October 16-19, 2005, accepted for publication.

\item{[3]} C. W. Wu and L. O. Chua, ``Synchronization in an array
of linearly coupled dynamical systems," IEEE Trans. CAS-I {\bf
42}, 430--447, 1995.

\item{[4]} X. F. Wang and G. Chen, ``Synchronization in
scale-free dynamical networks: robustness and fragility," IEEE
Trans. CAS-I {\bf 49}, 54--62, 2002.

\item{[5]} F. Vidyasagar, \emph{Nonlinear Systems Analysis},
Prentice-Hall, Englewood Cliffs, NJ, 1978.

\item{[6]} C. P. Li and X. Xia, ``On the bound of the Lyapunov exponents
for continuous systems", Chaos {\bf 14}, 557--561, 2004.

\item{[7]} C. P. Li and W. G. Sun, ``On synchronization of delayed
complex dynamical networks," The 24th Chinese Control Conference,
Guangzhou, July, 15-18, 2005, 202--205.

\item{[8]} W.H. Deng, Y.J. Wu and C.P. Li, ``Stability analysis
of differential equations with time-dependent delays," accepted
for publication in Int. J. Bifurc. Chaos {\bf 16}(2) (2006).

\item{[9]} G. Rangarajan and M. Z. Ding, ``Stability of synchronized
chaos in coupled dynamical systems," Phys. Lett. A {\bf 296},
204--209, 2002.

\item{[10]} C. P. Li and G. Chen, ``Estimating the Lyapunov
exponents of discrete systems," Chaos {\bf 14}, 343--346, 2004.

\item{[11]} C. G. Li and G. Chen, ``Synchronization in general
complex dynamical networks with coupling delays," Phys. A {\bf
343}, 263--278, 2004.

\end{document}